# RTK-Spec TRON:
# A Simulation Model of an ITRON Based RTOS Kernel in SystemC

M. AbdElSalam Hassan    Keishi Sakanushi    Yoshinori Takeuchi    Masaharu Imai

*Graduate School of Information Science and Technology, Osaka University, Osaka, Japan*
{hassan, sakanusi, takeuchi, imai}@ist.osaka-u.ac.jp

**Abstract**

*This paper presents the methodology and the modeling constructs we have developed to capture the real time aspects of RTOS simulation models in a System Level Design Language (SLDL) like SystemC. We describe these constructs and show how they are used to build a simulation model of an RTOS kernel targeting the μ-ITRON OS specification standard.*

## 1. Introduction

In recent years, as the embedded S/W started to increase in size and complexity; including firmware as device drivers, kernels, and boot code. An extra overhead started to appear from S/W side and it became time consuming to execute all this amount of S/W on simulated H/W using an instruction set simulator (ISS). So it was obvious that we need some higher level of co-simulation abstraction that can accelerate the co-simulation session and still maintain global system synchronization. This abstraction – referred here as *RTOS level simulation* is to simulate the embedded S/W consisting of the kernel and the application running on the top directly with H/W at the "C" source code level. This abstraction can be regarded as an evolution of host code execution technique, where the embedded S/W is not cross compiled to target code but compiled and executed as host code and its timing information is estimated a priori and annotated. After the emergence of SLDLs as SpecC [1] & SystemC [2], there have appeared some research results that are considering this level of abstraction as a candidate for simulating embedded S/W in a system level design methodology. The enabler of this new simulation abstraction is to accurately model and simulate an RTOS at the system level.

Researchers addressed this topic in variety of ways, using their own simulation engine as in [3,4], or SLDLs as in [5,6,7,8]. The differences mainly were in (i) the *design of the kernel simulation model*, whether they used an abstract generic RTOS and adopted a refinement methodology [3,6,7] or used the exact kernel code in simulation [4,5,8], (ii) the way to achieve *global system synchronization,* using a co-simulation manager [4] or a synchronization function [5], (iii) the modeling of *multi-tasking feature* of the application S/W; whether they used the SLDL simulation environment to perform context-switching [6] or used the multi-threading functionality of the host OS [5]. In our work, which is based on SystemC, we define two problems to be solved. First, although each kernel has its own specification that should be captured and modeled at the system level, in a market where over 40% of RTOSs are based on one specification standard i.e. *μ-ITRON* [9], our solution should also move towards a *System ITRON* standard. Second, the SystemC core language doesn't have the semantics to support preemption, thread priority assignment or scheduling necessary for RTOS modeling, so we should focus on extending the functionality of the language with *libraries* and *programming constructs* with *well defined semantics* that are capable of modeling embedded S/W performance.

Our contribution is that we present a novel technique to model and simulate *μ*-ITRON based RTOS kernels in SystemC. The technique depends mainly on isolating all *μ*-ITRON related dynamics in one simulation library and providing the designer with programming constructs that can be used to build a simulation model from the exact kernel implementation. We also propose a controllable process model, with execution semantics that support interruption, preemption, and further capable of gathering performance statistics including execution time and energy. To demonstrate the effectiveness of our approach, we build RTK-Spec TRON, a simulation model of the T-Kernel/OS [10], the core of the T-Engine system; an open development platform for embedded systems widely used in Japanese industry.

The rest of this paper is organized as follows. In section 2, we give an overview of RTK-Spec TRON. In section 3 and 4 we describe the RTOS modeling constructs we used to build the kernel simulation model, namely T-THREAD process and SIM_API library. In section 5, we present a case study on building a co-simulation framework based on RTK-Spec TRON and show different performance measures that can be gathered when running a video-game application in this framework. Section 6 concludes this paper.

## 2. Overview of RTK-Spec TRON

RTK-Spec TRON consists of three components: (i) T-Kernel/OS is the core that provides scheduling, resources, and resource management, (ii) T-Kernel/DS, acts as a debugger that references different resources and kernel internal states, and (iii) the user application, a programmable module wrapped in T-THREADs.

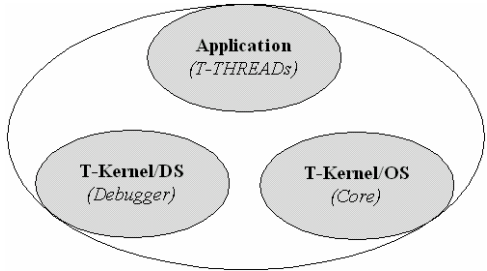

**Figure 1. RTK-Spec TRON Structure**

The T-kernel/OS is a real time OS that inherits ITRON technology, and further strengthens it. It employs a priority-based preemptive scheduling policy and supports several synchronization and communication mechanisms, including event flags, semaphores, mutexes, message buffers, and mailboxes. It provides a group of APIs for managing tasks, dynamic memory allocation (fixed and variable size pools), managing time (system time, cyclic, and alarm handling), interrupt handling, and system management. More details on T-Kernel and T-Engine system S/W can be found in [10].

## 3. T-THREAD Process

A Task Thread or shortly a T-THREAD process as shown in figure 2 was proposed here to capture the real time aspects of an application task or a handler (*cyclic, alarm, or external interrupt*) in embedded S/W. A T-THREAD is based on SystemC SC_(C)THREAD process [2] running under the supervision of a simulation API library (SIM_API) to simulate the behavior of a synchronized Petri-Net (PN) [11]. We used the following PN properties in implementation as execution semantics for the T-THREAD process model:

- A T-THREAD is a cyclic object of atomic transitions $T$ with a single token $K$ marking the state of the T-THREAD $m$. Only the activated state can fire if the corresponding enabling event occurs.
- An event $E$ that can occur within a T-THREAD belongs to one of RTOS kernel specific events $\Rightarrow E \in \{E_s, E_c, E_x, E_i, E_w\}$, $E_s$ is a startup event after kernel initialization, it is always associated with a source transition $T_o$. $E_c$ is a continue-run event denoting normal T-THREAD operation similar to an SC_(C)THREAD process. $E_x$ is a return from preemption event, when a given T-THREAD is preempted by a higher priority T-THREAD. $E_i$ is a return from an interrupt event when a T-THREAD is interrupted by another T-THREAD. $E_w$ is an arrival of a sleep event, when a T-THREAD voluntarily sleeps waiting for this event.
- Transitions are mapped to events based on the context at which the T-THREAD is executing, i.e. at startup, or within a service call, an application task, a handler, or H/W (BFM) access.
- A firing sequence $S$ is a sequence of transitions that is synchronized with the global system clock and causes a T-THREAD execution (*token moving*) from an initial marking to a designated marking. The execution time (delay) model associated to this firing sequence is denoted by $ETM(S|_{T-THREAD})$. The execution energy model associated to this firing sequence is denoted by $EEM(S|_{T-THREAD})$.
- A firing sequence $S$ has a characteristic vector $\underline{S}$ whose $i_{th}$ component is the number of times $T_i$ is fired in a firing sequence $S$.
- The consumed execution time/energy associated to a given place (marking) is denoted by $CET(\underline{S}|_{T-THREAD})$ and $CEE(\underline{S}|_{T-THREAD})$ respectively. These are accumulation of execution time/energy over multiple cycles of T-THREAD execution.

$$\Rightarrow CET(\underline{S}|_{T-THREAD}) = \sum_{Cycle=1}^{N} ETM(S|_{T-THREAD})$$

$$\Rightarrow CEE(\underline{S}|_{T-THREAD}) = \sum_{Cycle=1}^{N} EEM(S|_{T-THREAD})$$

- A *Scenario* is the execution sequence of different T-THREADs running on the top of a given kernel using a selected scheduler and under pre-determined timing constraints.

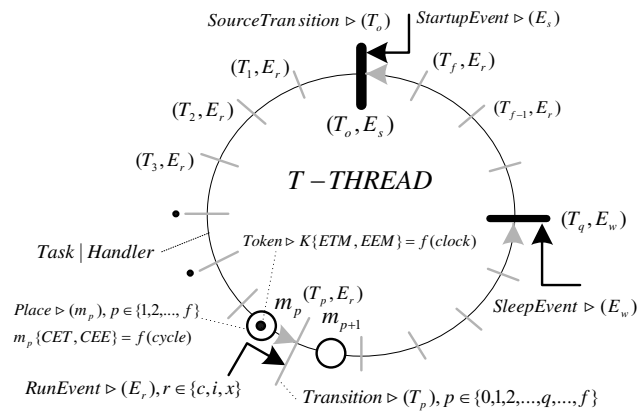

**Figure 2. T-THREAD Process Model**

## 4. SIM_API Library

To realize the T-THREAD process model, we extended SystemC simulation engine with a new simulation library represented by the APIs of Table 1. These APIs will be used as *programming constructs* from *the different modules* of an RTOS kernel simulation model to control the T-THREADs operation. The simulation dynamics of these APIs uses the $\mu$-ITRON v.4 specification standard [9] as a reference in kernel dynamics and depends on *events and dynamic sensitivity* feature introduced in SystemC 2.0 in implementation, therefore, building on an existing functionality in SystemC. Current dynamics support dispatching, delayed dispatching, service call atomicity[1], preemption - with system clock simulation granularity -, interrupts, and nested interrupts handling. The library contains a Thread hash table (SIM_HashTB) that keeps a record on every T-THREAD created upon startup and gets updated whenever a T-THREAD changes its state, a stack (SIM_Stack) data structure to model nested interrupts, and it interacts directly with external schedulers to schedule the next T-THREAD to run. It also has a debugging option for displaying time GANTT chart, and energy statistics for all registered T-THREADs.

To guarantee SIM_API coverage to capture real RTOS dynamics, we used SIM_API to build three kernel simulation models: *RTK-Spec I, II, and TRON*. RTK-Spec I (round robin scheduler) [8] and II (priority-based preemptive scheduler), are examples of user defined kernel specifications running on 8051 micro-controllers, and RTK-Spec TRON is an example of industrial kernel specification, that is widely used in Japan.

Figure 3 gives an illustration of RTK-Spec TRON simulation dynamics and SIM_API usage. In the shown dynamics, the kernel simulation model consists of a central module having three SC_THREADs: Thread Dispatch, Interrupt Dispatch and Boot Modules sensitive to s*ystem tick, external interrupts, and reset signals* respectively. Thread Dispatch activates the timer handler inside the T-Kernel/OS. The timer handler updates the system clock, checks for cyclic, alarm events, or task resuming events in the timer queue, it then calls simulation library APIs to start running a task/handler or preempt the running task if a task of higher priority is ready to run. Interrupt Dispatch identifies and responds to external interrupts by calling a simulation API to notify, their dedicated interrupt service routines. Boot is responsible for kernel startup sequence upon receiving H/W reset, i.e. initializing the kernel internal state and starting the initialization task, that will consequently call the user main entry to create & start tasks, handlers and allocate application resources.

T-Kernel/OS wait services like tk_slp_tsk will call a simulation API to indicate that a given task is waiting for a sleep event and request for context switching to a new task based on T-Kernel/OS scheduling policy. The waiting task will be notified later, upon the arrival of its event.

Each task or a handler will be assigned to a T-THREAD and a token gathers execution time/energy statistics as it propagates through different T-THREADs. To enable interruption and preemption; SIM_Wait will be used in a T-THREAD. SIM_Wait inherits *sc_wait* capability to model execution time and extends it to model energy. Furthermore, checking of *interruption* or *preemption* will be performed within SIM_Wait and when a T-THREAD is interrupted or preempted, it is scheduled to suspend upon reaching the next preemption point.

**Table 1. RTOS Modeling APIs** *(partial)*

| | |
|---|---|
| SIM_(Un)Register | Register/Delete a T-THREAD into/from T-THREAD Hash Table. |
| SIM_Init | Initialize T-THREAD Hash Table for a given number of T-THREADs. |
| SIM_AllCreated | Inform that all T-THREADs were created. |
| SIM_ChkAllCreated | Check if all T-THREADs were created. |
| SIM_WaitRun | Stop the execution of a T-THREAD waiting for $E_r$ event notification. |
| SIM_WaitEvent | Stop the execution of a T-THREAD waiting for $E_w$ event notification. |
| SIM_Wait | Stop the execution of a T-THREAD waiting for $E_y$ event notification. *If not interrupted or preempted, propagate token for ETM/EEM unit* |
| SIM_Start | Start the execution of a T-THREAD. |
| SIM_Stop | Stop the execution of a T-THREAD. |
| SIM_Schedule | Schedule for next T-THREAD to run without notification to start. |
| SIM_ContextSwitch | Schedule for next T-THREAD, suspend the current & notify the next one. |
| SIM_Preempt | Signal that a given T-THREAD was preempted. |
| SIM_Interrupt | Signal that a given T-THREAD was interrupted. |
| SIM_RetInt | Return from a Handler *(Cyclic, Alarm or External Interrupt)*. |
| SIM_StrtCritical | Inform the beginning of a Critical Section. |
| SIM_EndCritical | Inform the end of a Critical Section. |
| SIM_ChkCritical | Check if a Critical Section is executing. |
| SIM_Running | Inform that a given T-THREAD is running. |
| SIM_Sleeping | Inform that a given T-THREAD is sleeping. |
| SIM_Deferring | Inform that a given T-THREAD is deferring *(scheduled)*. |
| SIM_GUIinit | Interface SIM_API with Debugging & Performance Evaluation GUI Widgets. |

## 5. Case Study

In this section, we present a case study, on building an RTOS centric co-simulation framework based on RTK-Spec TRON. Our co-simulation framework example consists of RTK-Spec TRON, bus functional model (BFM) of i8051 MCU – modeled at register transfer level (RTL), a group of ASIC components connected to the BFM and wrapped in GUI widgets to give the look & feel of a virtual system prototype, and a video game application. An overall picture of the co-simulation framework is shown in figure 5. A description of the BFM and the application tasks module is presented next.

---
[1] Delayed Dispatching - A preemption that takes place within an interrupt handler or a nested interrupt handler is postponed till after the interrupt handler returns. Service Call Atomicity - All system calls issued by the user are executed with continuity.

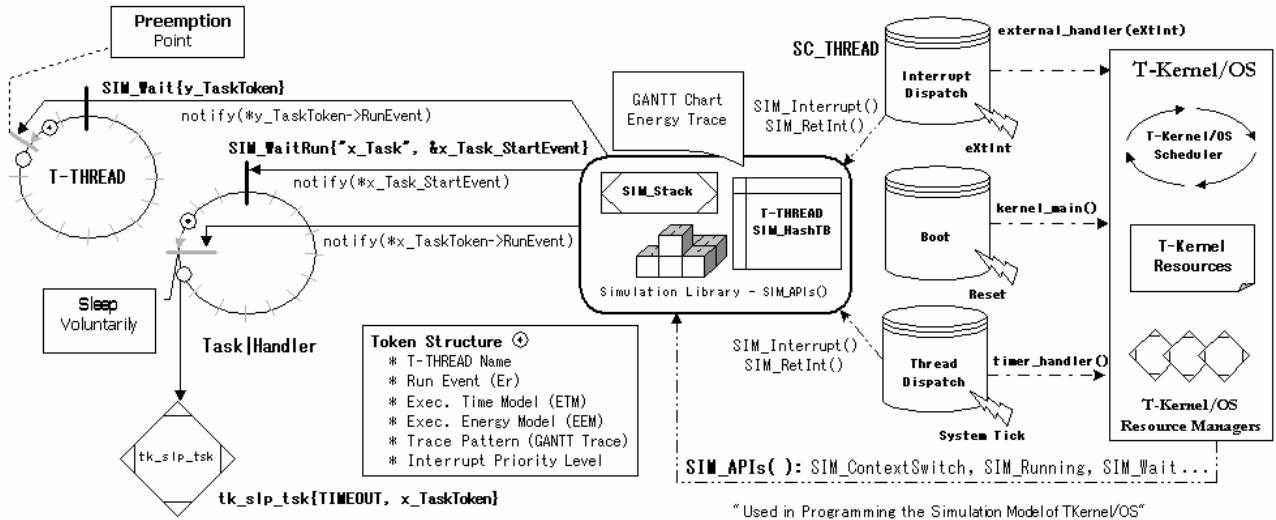

**Figure 3. Kernel Dynamics & SIM_API Usage**

### 5.1. Bus Functional Model (BFM)

A bus functional model is a key component in an RTOS centric co-simulation framework. It is an abstraction that models the external behavior of a processor with the surrounding H/W. Modeling can be either at transaction level (TLM) or register transfer level (RTL). For our experiment, we modeled a cycle accurate bus functional model that approaches the 8051core architecture in many structure and timing aspects at register transfer level. Interface was simplified by using SystemC as a common modeling platform. It is based on a Driver Model (*handshake functions)*, and represented by BFM calls as shown in figure 4. Each BFM Call will be associated with a cycle budget that is based on BFM timing characteristics, and an estimation on the energy consumed during that BFM access. As detailed in figure 5, the BFM consists of: *Real Time Clock* driving the kernel Central Module with default timing resolution = 1 ms, *Memory controller, Interrupt controller, Serial I/O*, and *Multiplexed Parallel I/O* interface to which several external peripheral devices are connected.

### 5.2. Application Tasks Module

This module represents the programmable entry for the user. Within each task, the user is able to access T-Kernel/OS services, e.g., waiting for a message or signal a semaphore and access member functions of the BFM to interact with H/W peripherals, e.g., reading from a memory location or writing to an I/O port or a serial buffer.

In our experiment here, *we programmed* a video game application that maps into four communicating tasks: {LCD:T1, Key pad:T2, SSD:T3, IDLE:T4} and two handlers {Cyclic:H1, Alarm:H2}, as shown in figure 4.

To measure the co-simulation speed of the overall framework including the overhead of GUI, the proposed modeling constructs, and SIM_API dynamics, we simulated the overall system for 1 s as a reference unit time S and measured the wall clock time R, considering different BFM access rates driving the GUI widgets – *wrapping the H/W peripherals* and different adjustments of the host CPU clock that avoids GUI display hazards. Simulation data in table 2, showed us that co-simulation speed (R/S) was lagging by 5X (S/R = 0.2) from real time without GUI overhead and 10X (S/R = 0.1) with GUI overhead and maximum BFM access driving a GUI widget every 10 ms. This speed was fast enough to display an animation on the LCD widget and capture user events. However, it would be difficult to animate and watch a display of GANTT chart or signal waveform changing at run time, as simulation will be overloaded with GUI callback functions and writing waveform to a log file. The simulation has to be performed in step mode, meaning that we advance simulation in step of system tick (1ms), rather than animate mode. Simulation was carried out on Pentium III 1.4 GHz/9% CPU average load.

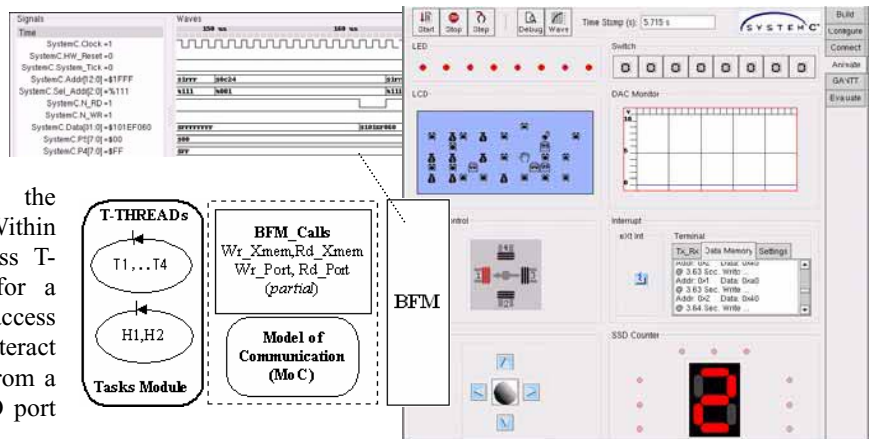

**Figure 4. Interaction with BFM–H/W Peripherals**

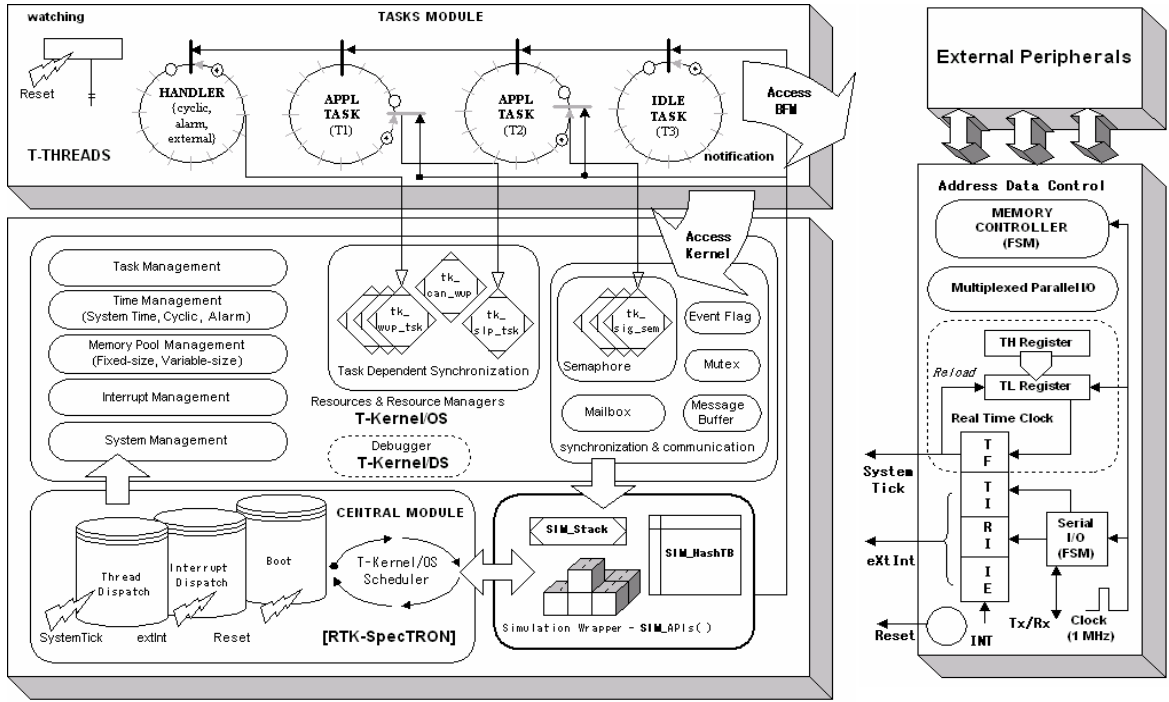

**Figure 5. RTOS Centric Co-Simulator in an SLDL**

Our conclusion was that; performing simulation at RTOS level; significant speed gain can be obtained compared to the RTL or ISS level co-simulation measures reported in [12]. This is mainly because of (i) the host code execution nature of *RTOS level simulation,* (ii) SystemC cycle based simulation, and (iii) using the same language to model the S/W, H/W, and GUI simplifying communication to direct C++ calls and removing the overhead of inter-process communication if different modeling languages where used instead.

**Table 2. Co-Simulation Speed Measure**

| Simulation Time (S) | 1 s | | | |
|---|---|---|---|---|
| BFM Access Rate (every) | No Access | 10 ms (maximum) | 1 ms | 1 µs |
| Wall Clock Time (R) | 5 sec | 10 sec | 16 min | 50 min |
| GUI Refresh Rate (Host CPU Clock) | No GUI | 10 ms | 1000 ms | 3000 ms |
| Speed Gain (S/R) | 0.2 | 0.1 | 0.001 | ~0(No gain) |
| Speed Lag (R/S) | 5X | 10X | 960X | 3000X |
| Mode | Console | Animate | Step | Step |
| Example | Send Peripherals Data To Log File | LCD/MPEG Animation | System Tick Tracing | GANTT Chart & Waveform Tracing |

We also developed a group of GUI widgets that allow users debug and optimize their application S/W in terms of processing performance and power dissipation. Results can be displayed in different ways. The most interesting diagrams are, (i) Execution Time/Energy Trace widget – figure 6 (available in *step mode*). In this widget, task dispatching, interrupt handling, and preemption can be observed. Also, different contexts of execution are assigned different patterns to display the execution time/energy of a BFM access, basic block, or OS service. (ii) Time/Energy distribution widget – figure 7 (available in *animate mode*). In this widget, a battery of 10-watt-hour was assumed and at run time the consumed execution time (CET) and energy (CEE) were accumulated and distributed over registered T-THREADs and the battery's status bar was updated. From such a display, designers can figure out the maximum duration of the battery's lifespan for a given application, and the tasks that consume much time or energy, hence making good decision for HW/SW partitioning, by moving some S/W tasks to H/W or optimizing tasks' code. Other debugging widgets are tracing T-kernel internal states and resource usage using T-Kernel/DS functions as shown in figure 8, and monitoring H/W by probing signals and variables in a waveform viewer as was previously shown in figure 4.

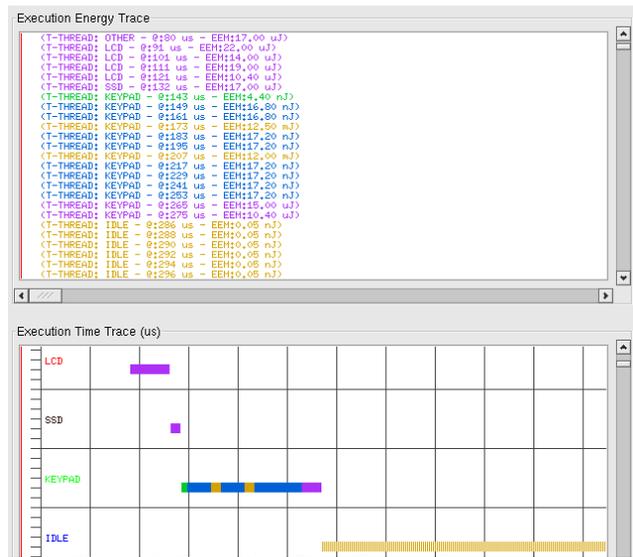

**Figure 6. Execution Time/Energy Trace** *(colored)*

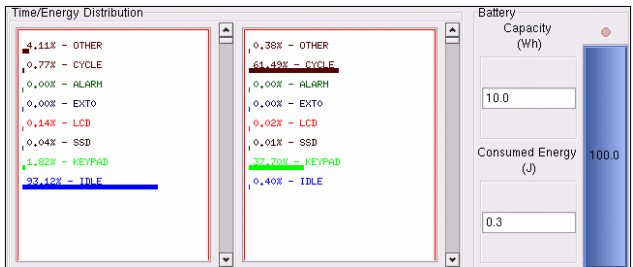

**Figure 7. Consumed Time/Energy Distribution**

Estimation of both execution time and energy of the kernel & application tasks at the "C" source code level plays a crucial role in RTOS level simulation abstraction. By accurate *calibration* of the kernel simulation model and good *estimation* of the tasks runtime; the co-simulation accuracy can be relatively high compared to ISS co-simulation as we reported in [8]. In this case study however, the ETM/EEM annotations we used for RTK-Spec TRON and the application tasks were estimated. By cross profiling or calibration against ISS or T-Engine emulation, for a given supported T-Engine platform based architecture, we can raise the accuracy of co-simulation, and create a virtual prototype of the application running on the synthesis platform (OS+H/W). An investigation of this issue is the subject of our future study.

**Figure 8. T-Kernel/DS Output Listing** *(sample)*

## 6. Conclusion

Increasing software content & design complexity are driving designers to consider new methodologies of simulating their systems. *RTOS level simulation* is one candidate that fits seamlessly with SLDL & moves in the direction of raising the modeling abstraction; independent on the underlying processor architecture or instruction set used. We addressed the problem of simulating RTOS kernel implementations that follow the $\mu$-ITRON specification standard. Our solution proposed programming constructs & library support in SystemC & created a simulation model of one kernel implementation that inherits ITRON technology, i.e. T-Kernel/OS. Finally, we built an RTOS centric co-simulation framework based on the simulated model & showed different performance measures that can be gathered & help designers develop & test their embedded systems early at design cycle.